\documentclass[12p,a4paper]{article}
\usepackage{epsfig}
\usepackage{amssymb}
\begin{document}

\title{Study of the nucleon spin-dependent structure function $g_1$.\\
A comparison with recent HERMES and COMPASS data.}

\author{Dorota Kotlorz\footnote{Opole University of Technology, Division of
Physics, Ozimska 75, 45-370 Opole, Poland, e-mail: 
{\tt d.strozik-kotlorz@po.opole.pl}} and Andrzej Kotlorz\footnote{Opole
University of Technology, Division of Mathematics, Luboszycka 3, 45-036 Opole,
Poland, e-mail: {\tt a.kotlorz@po.opole.pl}}}
\date{April 1, 2008}
\maketitle

\abstract{
Predictions for the spin dependent structure function $g_1$ of the nucleon
are presented. We use an unified approach incorporating the LO DGLAP
evolution and the resummation of double logarithmic terms $ln^2(x)$.
We show, that the singular input parametrisation as $x\rightarrow 0$
can be a substitute of the $ln^2(x)$ resummation. An impact of the
`more running' coupling is discussed. We determine the contribution to
the Bjorken sum rule solving the evolution equation for the truncated
moment of $g_1^{NS}$. A comparison with the re-analysed HERMES and COMPASS
data is given.\\\\
PACS{12.38.Bx}
}

\section{Introduction}
\label{Intro}

Experimental data confirm (at least for $Q^2>1$ ${\rm GeV}^2$) the theoretical
predictions of an increase of the nucleon structure functions at small values
of the Bjorken $x$.
The low-$x$ behaviour of both spin averaged and spin dependent structure
functions is controlled by the double logarithmic terms $(\alpha_s ln^2(x))^n$
\cite{b1}-\cite{b3}. In a unpolarised case, this singular PQCD behaviour is
however overridden by the leading Regge contribution present in the input
parametrisation \cite{b4}.
The situation is quite different in the spin-dependent case, where
the double logarithmic effects are very important. The resummation of the
$ln^2(x)$ terms at low $x$ goes beyond the standard LO and NLO PQCD evolution
of the parton densities. Double logarithmic contributions become essential for
$x\sim 0.01$, where there is little experimental data. Determination of the
sum rules and the nucleon spin decomposition among partons requires knowledge
of the structure functions over the entire region of the variable $x\in (0;1)$.
Therefore the small-$x$ behaviour of the spin dependent parton distributions
is a topic of the intensive theoretical investigations. Standard approach
describing structure functions is based on the DGLAP-$Q^2$ evolution equation
via two-step convolution: of the initial parton densities and splitting
functions and then of the evolved parton distributions and the coefficient
functions. Because there is no way to calculate the initial parton densities,
which have a nonperturbative origin, they must be put `by hand'. Different
parametrisations of the initial gluon and quark densities known in literature
e.g. \cite{b5,b6} are singular when $x\rightarrow 0$. This choice enables to
study DIS phenomena within DGLAP approach not only for the large-$x$ region
but for the small-$x$ one as well. Singular terms $\sim x^{-\lambda}$ can be
a substitute of the double logarithmic $ln^2(x)$ resummation, which is absent
in the standard DGLAP scenario. This problem has been widely discussed and
argued in \cite{b7}. Also important is the problem of the $\alpha_s$ dependence
of the QCD evolution. Following \cite{b8} we take into account the running
coupling effects not via $\alpha_s (Q^2)$ but with use of the more running
$\alpha_s (Q^2/x)$. This approach is better justified at small values of $x$,
whereas for large $x\sim 1$ leads to the usual DGLAP coupling $\alpha_s (Q^2)$.
It seems to be reasonable to study an impact of the double logarithmic and
running coupling effects on theoretical predictions for spin structure
functions.

In this paper we present the unified approach, in which the familiar $Q^2$
evolution is extended by the $ln^2(x)$ resummation. In our analysis we use
so-called unintegrated parton distributions and solve the combined LO DGLAP
$+ ln^2(x)$ evolution equation with help of the Chebyshev polynomial technique.
We take into account the `very running' coupling effects at small $x$ and
discuss the role, they play.
We also show that the singular input parametrisation of the parton
distributions can be some kind of substitution for the double logarithmic
terms, missing in the standard DGLAP approximation.
Our theoretical predictions for the spin dependent structure function $g_1$
are compared with recently re-analysed HERMES and COMPASS data.

The content of this paper is as follows. In Section 2 
we recall the unified approach incorporating DGLAP evolution of
structure functions and the double logarithmic $\alpha_s^n ln^{2n}(x)$
terms, which are essential in the small-$x$ region.
Section 3 is devoted to the impact of running coupling effects
on the $g_1$ results in the small-$x$ region. 
In Section 4 we show that the singular initial parton densities
$\sim x^{-\lambda}$ can mimic the resummation of double logarithmic terms
$ln^2(x)$. 
Using this fact, in Section 5 we solve the evolution equation
for the truncated moments themselves and obtain contribution to the Bjorken
sum rule. We also present numerical predictions for the structure
function $g_1$ and compare them to re-analysed HERMES and COMPASS data.
Finally, in Section 6 we summarise our results.

\section{Unified $ln^2(x)+$LO DGLAP approach}
\label{sec.2}

The structure functions of the nucleon can be expressed in terms of the parton
distributions. These depend on two kinematic variables: the Bjorken $x$ and
$Q^2=-q^2$ with $q$ being the four-momentum transfer in the deep-inelastic
lepton-nucleon scattering (DIS). The scaling variable is defined as
$x=Q^2/(2pq)$,
where $p$ is the nucleon four-momentum. The strong interactions between quarks
and gluons cause the changes in the parton densities. For medium and large $x$,
the evolution with $Q^2$ of the parton distributions is well described by
the standard Dokshitzer-Gribov-Lipatov-Altarelli-Parisi (DGLAP) equations
\cite{b9}-\cite{b12}. This approach which effectively sums up the leading
$ln(Q^2)$ terms is however incomplete at small $x$, where another large
logarithm - $ln(1/x)$ becomes essential and which leading powers
$\alpha_s^n ln^{2n}(x)$ needs to be resummed. The double logarithmic terms
$ln^2(x)$ come from the ladder diagrams with quark and gluon exchanges along
the chain. Treating both potentially large logarithms $ln(Q^2)$ and $ln(1/x)$
on equal footing, the authors of \cite{b13}-\cite{b15} obtained equations which
incorporate DGLAP evolution and $ln^2(x)$ terms as well. The double logarithmic
effects go beyond the standard LO and even NLO $Q^2$ evolution of the
spin dependent parton distributions and significantly modify the Regge
pole model expectations for the structure functions. Theoretical analyses
of the small-$x$ behaviour of the polarised structure functions
\cite{b16} predict that resummation of the double logarithmic
terms $(\alpha_s ln^2(x))^n$ leads to the singular form as $x\rightarrow 0$:
\begin{equation}\label{r2.1}
g_1^{NS,S}(x,Q^2) \sim x^{-\lambda_{NS,S}}\;,
\end{equation}
where $\lambda_{NS}\approx 0.4$, $\lambda_{S}\approx 0.8$ and $g_1^{NS,S}$
denotes nonsinglet or singlet part of the polarised structure function of
the proton.
For larger but still low $x\in(10^{-5};10^{-2})$, $g_1$ is less steep with
the slope $\lambda\approx 0.2-0.3$ for the nonsinglet part \cite{b1}.
This power-like behaviour $x^{-\lambda}$ remains significantly steeper
than the DGLAP solution in absence of the singular input parametrisations
of parton densities
\begin{equation}\label{r2.2}
g_1^{DGLAP}(x\rightarrow 0)
\sim \exp\sqrt{\ln(1/x)\ln\ln(Q^2/\Lambda_{QCD}^2)}\;.
\end{equation}
The unified equation, which includes the LO DGLAP evolution and $ln^2(x)$
terms resummation reads:
\begin{eqnarray}\label{r2.3}
f(x,Q^2)&=&f_{0}(x)+
\underbrace{\int\limits_{Q_0^2}^{Q^2}\frac{dk'^2}{k'^2}\;\frac{\alpha_s}{2\pi}\;
\Delta P \otimes f(x,k'^2)}_{DGLAP}\nonumber\\
&+&\underbrace{\frac{4}{3}\int\limits_x^1\frac{dz}{z}\int\limits_{Q^2}^{Q^2/z}
\frac{dk'^2}{k'^2}\;\frac{\alpha_s}{2\pi}\;
f\left(\frac{x}{z},k'^2\right)}_{LN^2(X)\:\:\:LADDER}\nonumber\\
&+&\underbrace{Bremsstrahlung\:\: corrections}_{LN^2(X)\:\:\: NONLADDER}\:\:,
\end{eqnarray}
where $\otimes$ abbreviates a Mellin convolution over $x$
\begin{equation}\label{r2.3a}
(\Delta P \otimes f)\:\:(x,Q^2) = \int\limits_{x}^{1}\frac{dy}{y}\;
\Delta P\left ( \frac{x}{y}\right )\; f(y,Q^2)\;,
\end{equation}
$\Delta P$ denote the polarised version of the splitting function $P$ and $f$
is the unintegrated distribution, related to the ordinary polarised parton
distribution $\Delta p(x,Q^2)$ via
\begin{equation}\label{r2.4}
f(x,Q^2)=\frac{\partial \Delta p(x,Q^2)}{\partial\ln(Q^2)}\;.
\end{equation}
The double logarithmic terms come from ladder-type graphs as well as from
the nonladder ones which represent radiative corrections \cite{b2},
\cite{b17}-\cite{b19}.
In a case of the nonsiglet polarised structure functions the contribution of
nonladder diagrams is negligible. However for the singlet spin dependent
structure functions, besides the ladder graphs, one has to include
Bremsstrahlung corrections \cite{b3}, which are important.
The full evolution equations for nonsinglet and singlet unintegrated parton
distributions within DGLAP$+ln^2 x$ approach have been presented in
\cite{b13,b14}. This forms the basis of our analysis in the next section,
where we discuss the modified running coupling effects at small $x$.

\section{Running coupling $\alpha_s$ in the small-$x$ region}
\label{sec.3}

DGLAP formalism uses the following prescription for the running coupling
(in the lowest order):
\begin{equation}\label{r3.1}
\alpha_s=\alpha_s(Q^2)=\frac{12\pi}{(33-2N_f)\:ln\frac{Q^2}{\Lambda_{QCD}^2}}\;,
\end{equation}
where $N_f$ is the number of active quark flavours and
$\Lambda_{QCD}\approx 200\;{\rm MeV}$ is
the QCD cut-off parameter. It has, however, been argued that in the small-$x$
region eq.(\ref{r3.1}) should be rearranged into the following form \cite{b8}:
\begin{equation}\label{r3.2}
\alpha_s=\alpha_s(Q^2/z)
\end{equation}
with $z$ being the longitudinal momentum fraction of a parent parton,
carried by a next generation parton. In this way $\alpha_s$
becomes `very running' i.e. runs in each ladder rung depending on the
gluon virtuality. This prescription of $\alpha_s$, widely discussed also
in \cite{b20,b21}, has been used e.g. in \cite{b1,b22}
within double logarithmic effect $ln^2(x)$ resummation. Here, we study
an impact of the different $\alpha_s$ parametrisation on the polarised
parton densities. Because
\begin{equation}\label{r3.3}
\alpha_s(Q^2/z)\leq\alpha_s(Q^2)\;,
\end{equation}
in the case of the `very running' $\alpha_s$ (\ref{r3.2}),
the growth of the parton distributions in the low-$x$ region is damped.
A scale of the damping for nonsinglet and singlet
(gluons) distributions is shown in Figs. 1-2, where we plot the ratio
\begin{equation}\label{r3.4}
R = \frac{\Delta p(\alpha_s(Q^2))}{\Delta p(\alpha_s(Q^2/z))}
\end{equation}
as a function of $x$. Here, $\Delta p$ denotes the nonsinglet (valence)
$\Delta q^{NS}$ and the gluon $\Delta G$ distribution function respectively.
\begin{figure}[ht]
\begin{center}
\includegraphics[width=90mm]{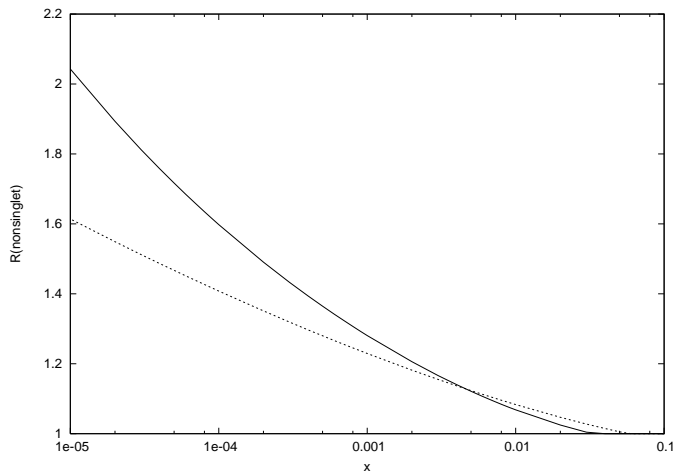}
\caption{The ratio (\ref{r3.4}) for the polarised nonsinglet quark distribution
$\Delta q^{NS}=\Delta u - \Delta d$ as a function of $x$ at
$Q^2=10$ ${\rm GeV}^2$.
Solid: unified DGLAP+$ln^2(x)$ approach, dotted: DGLAP alone.}
\end{center}
\end{figure}
%
\begin{figure}[ht]
\begin{center}
\includegraphics[width=90mm]{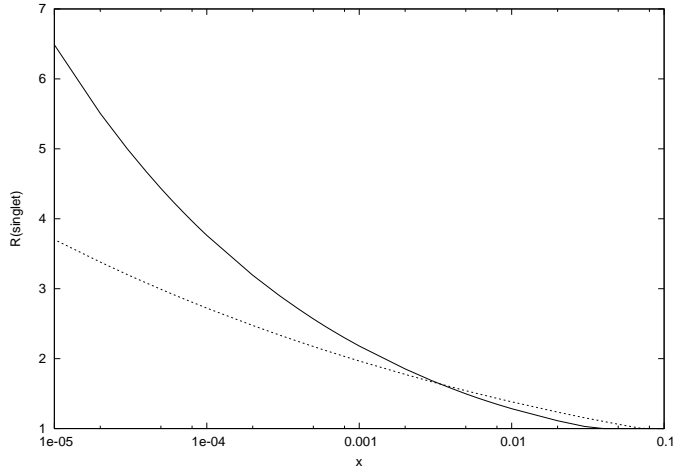}
\caption{The ratio (\ref{r3.4}) for the polarised gluon distribution
$\Delta G$ as a function of $x$ at $Q^2=10$ ${\rm GeV}^2$.
Solid: unified DGLAP+$ln^2(x)$ approach, dotted: DGLAP alone.}
\end{center}
\end{figure}
One can see, that the difference becomes essential at $x\sim 0.01$ and the
impact of the running coupling effects for the singlet case is larger than for
the nonsinglet one.
Double logarithmic resummation additionally amplifies the split between
results in comparison to the pure DGLAP approach.
At very small $x=10^{-5}$ we find the ratio (\ref{r3.4}) about 2 for the
nonsinglet polarised distribution and above 6 for the polarised gluons. 
Our estimations of $R$ (\ref{r3.4}) show that for the small values of
Bjorken parameter $x\leq 10^{-2}$ the coupling $\alpha_s(Q^2)$ should be
replaced by $\alpha_s(Q^2/z)$. In standard DGLAP analysis, where rather
large-$x$ region is considered, this modification converts into
$\alpha_s(Q^2)$ ($z\sim 1$). Parametrisation of the coupling $\alpha_s$
is not the only crucial point in the low-$x$ analysis of structure functions.
Another problem are initial parton distributions at low $Q_0^2\sim 1$
${\rm GeV}^2$, which enter into the evolution equations. The behaviour
of the quark and gluon distributions at very small $x$ is mainly generated
by the double logarithmic $ln^2(x)$ effects. Therefore singular as
$x\rightarrow 0$ inputs $\sim x^{-a_1}$ seem to be needless in PQCD analysis,
unless one does not consider $ln^2(x)$ terms. Within standard DGLAP approach,
parametrisations in a form $\sim x^{-a_1}$ can be regarded as a substitute
of the missing double logarithmic effects resummation. In the next
section we discuss this problem in detail.

\section{Singular input parametrisations as an ersatz of the double
logarithmic terms $ln^2(x)$ resummation}
\label{sec.4}

According to the philosophy of DGLAP approach, structure functions
of the nucleon are a convolution of the coefficient functions and
the evolved parton distributions. In this formalism, the polarised structure
function $g_1(x,Q^2)$ for the proton is given by \cite{b12a}
\begin{eqnarray}\label{r4.0}
g_1^p(x,Q^2)&=& \frac{1}{2}\langle e^2\rangle\big[
C_{NS}\otimes\Delta q^{NS}(x,Q^2)\nonumber\\
&+& C_{S}\otimes\Delta q^{S}(x,Q^2)+2N_f\;C_G\otimes\Delta G(x,Q^2)\big]\;,
\end{eqnarray}
where
\begin{equation}\label{r4.0a}
\langle e^2\rangle = \frac{1}{N_f}\sum\limits_{i=1}^{N_f}e_i^2\;.
\end{equation}
Here, $e_i$ denotes the electric charge of the $i$ quark-flavour, $\Delta q^{NS}$,
$\Delta q^{S}$, $\Delta G$ are respectively the nonsinglet and singlet
quark and the gluon polarised densities (helicity distributions). The
coefficient functions $C_i$ are computed to a given order in $\alpha_s$.
PQCD evolution equations for the quarks and gluons distribution functions
need the nonperturbative input quantities at some initial scale $Q_0^2$.
These input parametrisations, fitted to the experimental data, together with
the suitable PQCD framework provide a satisfactory agreement of theory
and measurements. The standard theoretical investigation of deep-inelastic
scattering structure functions based on the DGLAP approach concerned
originally the region of large $x$ and large $Q^2$. In this way the parton
evolution with respect to $Q^2$ is taken into account, whereas the evolution
with respect to Bjorken $x$ is neglected. In the small-$x$ region,
logarithms of $x$ become also important and therefore must be taken into
account. Assuming singular as $x\rightarrow 0$ initial parton distributions
$\sim x^{-a_1}$ ($a_1>0$), one can obtain within standard DGLAP approach a
substitute of the double logarithmic $ln^2(x)$ resummation, which is essential
at low $x\ll 1$. However, if we take into account the double logarithmic terms
via the suitable kernel of the evolution equations, we do not need to use the
`artificial support' in a form of the singular initial parametrisations. An
impact of the input parton distributions on the final (after evolution)
results is large. This is shown in Fig. 3, where we plot the LO DGLAP
evolution from $Q_0^2=1$ ${\rm GeV}^2$ to $Q^2=10$ ${\rm GeV}^2$ of the
nonsinglet polarised structure function $\Delta q^{NS}$.
We test different input parametrisations of the general form:
\begin{equation}\label{r4.1}
\Delta q^{NS}(x,Q_0^2)\sim x^{-a_1}(1-x)\,^{a_2}.
\end{equation}
There is no doubt, that the small-$x$ behaviour of the parton densities
is dominated just by the $x^{-a_1}$ term, which survives the QCD evolution
when $a_1>0$.
\begin{figure}[ht]
\begin{center}
\includegraphics[width=90mm]{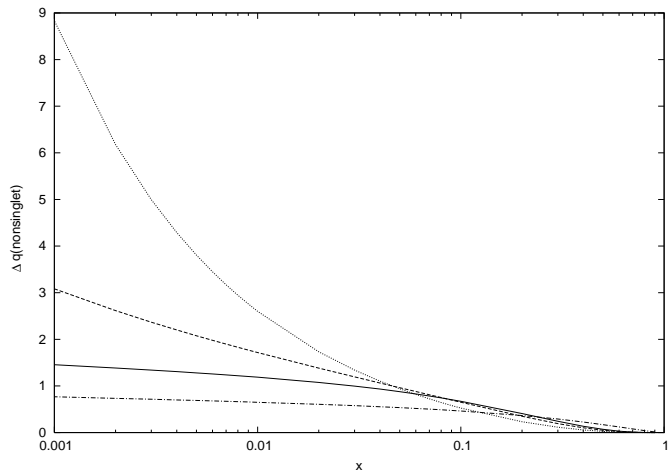}
\caption{The LO DGLAP evolution from $Q_0^2=1$ ${\rm GeV}^2$ to
$Q^2=10$ ${\rm GeV}^2$ of the nonsinglet polarised structure function
$\Delta q^{NS}$ as a function of $x$. Different input parametrisations
(\ref{r4.1}). Solid: $a_1=0$, $a_2=3$; dashed: $a_1=0.2$, $a_2=3$;
dashed-dotted: $a_1=0$, $a_2=1$; dotted: $a_1=0.5$, $a_2=3$.}
\end{center}
\end{figure}
Hence appropriate choice of the initial conditions must be consistent
with used theoretical treatment.
Thus there are two possible scenarios. Either we consider the unified
evolution equations with two parts of the kernel: the standard DGLAP one
and the other one - generating $ln^2(x)$ terms. Then the input distributions
are assumed to be nonsingular as $x\rightarrow 0$. In this case the small-$x$
behaviour of the structure functions is totally governed by the evolution.
Or we use the pure DGLAP analysis together with the singular parametrisations,
which mimic the missing at low-$x$ resummation of the leading logarithms.
In Fig. 4 we plot  the logarithm of the polarised nonsinglet and gluon
distributions evaluated at $Q^2=10$ ${\rm GeV}^2$ within unified
DGLAP$+ln^2(x)$ approach as a function of $ln(1/x)$. We can estimate the
effective slopes of the presented curves $\lambda (x,Q^2)$, defined as:
\begin{equation}\label{r4.1a}
\lambda_p (x,Q^2) = \frac{\partial\ln [\Delta p(x,Q^2)]}
{\partial\ln (\frac{1}{x})}\;.
\end{equation}
Here, $\Delta p$ denotes again respectively the nonsinglet quark and gluon
helicity distributions ($\Delta q^{NS}$, $\Delta G$).
From the plots we find, namely, $\lambda_{NS}\approx 0.2$ and
$\lambda_{G}\approx 0.6$. Hence the `ersatz' input $\sim x^{-\lambda}$,
which is able to reproduce the double logarithmic $ln^2(x)$ resummation in the
small-$x$ region $x\in (10^{-4}\; ;\; 10^{-2})$ should have a form:
\begin{equation}\label{r4.2}
\Delta q^{NS}\sim x^{-0.2}
\end{equation}
for the nonsinglet part and
\begin{equation}\label{r4.3}
\Delta G\sim x^{-0.6}
\end{equation}
for the gluons respectively.
\begin{figure}[ht]
\begin{center}
\includegraphics[width=90mm]{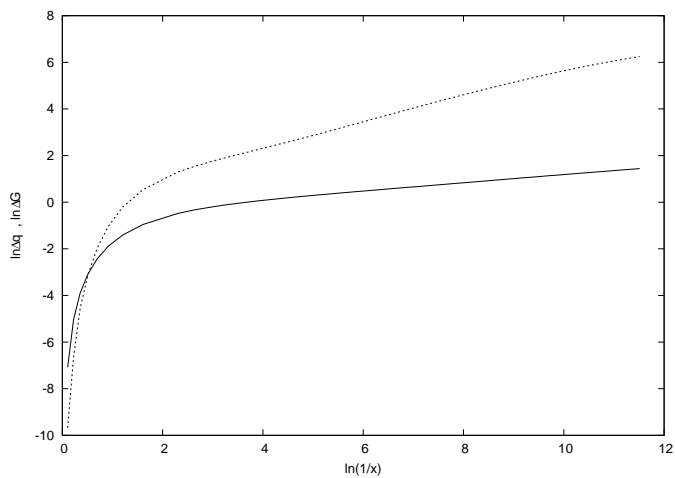}
\caption{The logarithm of the polarised quark nonsinglet (solid) and gluon
(dotted) distributions evaluated at $Q^2=10$ ${\rm GeV}^2$ within unified
DGLAP$+ln^2(x)$ approach as a function of $ln(1/x)$. An illustration of the
slope $\lambda$, defined in (\ref{r4.1a}).}
\end{center}
\end{figure}
As one can see, the behaviours (\ref{r4.2}), (\ref{r4.3}) are less steep than
their asymptotic limits as $x\rightarrow 0$:
\begin{equation}\label{r4.4}
\Delta q^{NS}(x\rightarrow 0)\sim g_1^{NS}(x\rightarrow 0)\sim x^{-0.4}
\end{equation}
and
\begin{equation}\label{r4.5}
\Delta G(x\rightarrow 0)\sim g_1^{S}(x\rightarrow 0)\sim x^{-0.8}\;.
\end{equation}
The results (\ref{r4.4}), (\ref{r4.5}) were obtained in \cite{b14} via
estimation of the anomalous dimensions and also in \cite{b16} - within IREE
(infrared evolution equation) formalism.

In conclusion, the power-like behaviour $x^{-\lambda}$ of the quark and
gluon polarised distribution functions, generated by the resummation of
the double logarithmic terms $ln^2(x)$, can be also obtained via the singular
factors in the initial parton distributions.
Finally, let us shortly discuss the possible values of $a_1$ in the leading
term of the initial parton densities. The choice of the value of $a_1$
in the input parametrisations (\ref{r4.1}), which controls the singular
small-$x$ behaviour, depends on the evolution length ($Q^2-Q_0^2$). If one
assumes a very low input scale $Q_0^2\lesssim 1$ ${\rm GeV}^2$, then
already the smaller value of $a_1=0.2$ in the nonsinglet case can `mimic' the
$ln^2(x)$ effects. In contrast, for longer $Q_0^2\approx 4$ ${\rm GeV}^2$, what
denotes the shorter evolution, one should use more singular input with
$a_1\approx 0.4$ for the nonsinglet case. Similar (or even more singular) input
parametrisations of the spin-dependent parton distributions have been assumed
e.g. in \cite{b5}, \cite{b6}, \cite{b30}.
In the next section we shall compare our theoretical predictions based on
either the unified DGLAP$+ln^2(x)$ approach or the DGLAP analysis alone
together with the singular inputs, with experimental data.

\section{Comparison with experimental data}
\label{sec.5}

In this section we shall present our results obtained using the unified
DGLAP$+ln^2(x)$ approach (\ref{r2.3}) with `very running' $\alpha_s$
(\ref{r3.2}). We shall also apply an alternative scenario, described in the
previous section, in which the pure DGLAP analysis is accompanied by the
singular input parton densities at the low scale $Q_0^2=1$ ${\rm GeV}^2$.
In this latter approach we shall compute i.a. the truncated Mellin moments
of structure functions using directly the evolution equations for truncated
moments, derived recently in \cite{b23}.
Let us recall now some basic formulas concerning this approach.

The evolution equations for the truncated moments of the parton densities
have the form:
\begin{equation}\label{r5.1}
\frac{d\bar{q}_n(x_0,Q^2)}{d\ln Q^2}=  
\frac{\alpha_s(Q^2)}{2\pi}\; (P'\otimes \bar{q}_n)(x_0,Q^2),
\end{equation}
\begin{equation}\label{r5.2}
P'(n,z)= z^n\, P(z),
\end{equation}
where $P(z)$ is the well-known splitting function from
the DGLAP equation.
$\bar{q}_{n}(x_0,Q^2)$ denotes the $n$th Mellin moment of the
distribution function $q(x,Q^2)$ truncated at $x_0$:
\begin{equation}\label{r5.3}
\bar{q}_{n}(x_0,Q^2)=\int\limits_{x_0}^1 dx\, x^{n-1}\, q(x,Q^2).
\end{equation}
This formula is obviously valid also in the spin-dependent case where one
replaces $q$ by $\Delta q$, $\bar{q}_n$ by $\Delta\bar{q}_n$ and
$P$ by $\Delta P$ e.i. the unpolarised quantities by their `polarised'
versions.
It is particularly interesting to note that the evolution equation
for the $n$th truncated moment has the same form as that for the parton
density function itself with the modified splitting function $P'$ (\ref{r5.2}).
The truncated moments approach refers directly to the physical values - moments
(rather than to the parton distributions), what enables one to use a wide range
of deep-inelastic scattering data in terms of smaller number of parameters.
In this way, no assumptions on the shape of parton distributions are needed.
Using the evolution equations for the truncated moments one can also avoid 
uncertainties from the unmeasurable very small $x\rightarrow 0$ region.
Eq. (\ref{r5.1}) does not account for the double logarithmic $ln^2(x)$ terms
resummation, what can be mimiced by the appropriate input, as it was
described in the previous section. This is the motivation that we use the
equations for the truncated moments in the studies presented here.

In order to compare theoretical predictions with experimental data over the
kinematic range explored one should generalize the results to the
small-$Q^2<1$ ${\rm GeV}^2$ region. Thus, one can use the prescription
introduced in \cite{b24} and applied in the studies \cite{b13,b14,b25,b26},
valid for arbitrary $Q^2$:
\begin{equation}\label{r5.3a}
Q^2\rightarrow Q^2+Q_0^2
\end{equation}
and
\begin{equation}\label{r5.3b}
x\rightarrow \bar{x}=(Q^2+Q_0^2)/(2pq).
\end{equation}
After this rearrangement the structure function $g_1$ can be extrapolated to
the low-$Q^2$ region (for fixed 2pq) including the point $Q^2=0$, although
perturbative $Q^2$-power and higher twist corrections may also play a role
in this region \cite{b27}. Taking into account the small-$Q^2$ corrections
is particularly important when one studies recent COMPASS measurements
obtained for very small $4\cdot 10^{-5}<x<2.5\cdot 10^{-2}$ at simultaneously
very low $Q^2\ll 1$ ${\rm GeV}^2$ \cite{b29a}.
We use in our analysis input parametrisations of polarised parton densities
at the initial scale $Q_0^2=1$ ${\rm GeV}^2$ in a simple general form:
\begin{equation}\label{r5.4}
\Delta q(x,Q_0^2) = \eta\: x^{-a_1}(1-x)^{a_2},
\end{equation}
where $\eta$ is a normalization factor. The exponent $-a_1$ controls
the behaviour of $\Delta q$ in the small-$x$ region and the factor
$(1-x)^{a_2}$ ensures the vanishing of the parton density as $x\rightarrow 1$.
The singular part $x^{-a_1}$, where $a_1>0$, can  mimic the resummation of
the leading logarithms $ln^2 x$.\\
\begin{figure}[ht]
\begin{center}
\includegraphics[width=90mm]{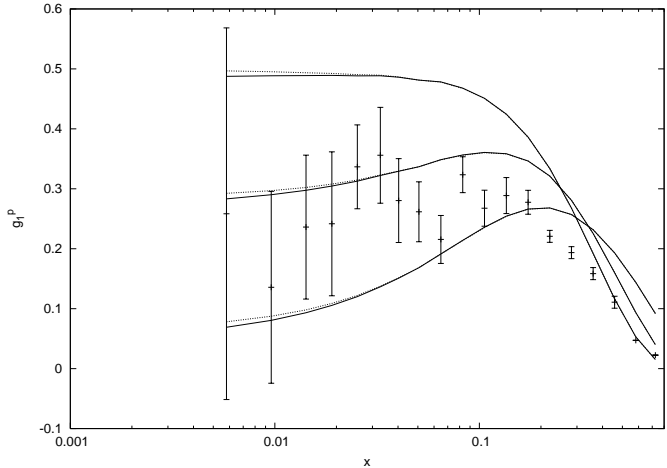}
\caption{The spin dependent structure function for proton $g_1^p$ vs $x$,
compared with HERMES data. $Q^2$ is the measured mean value 
$\langle Q^2 \rangle$ at each $x$.
Plots for different input parametrisations of the valence quarks
$\sim (1-x)^{a_2}$ from up to bottom at $x=0.01$: $a_2=3$, $a_2=2$,
$a_2=1$. Solid (dotted) line corresponds to the positive (negative) solutions
for polarised gluons $\Delta G$ . Error bars represent the statistical
uncertainties.}
\end{center}
\end{figure}
\begin{figure}[ht]
\begin{center}
\includegraphics[width=90mm]{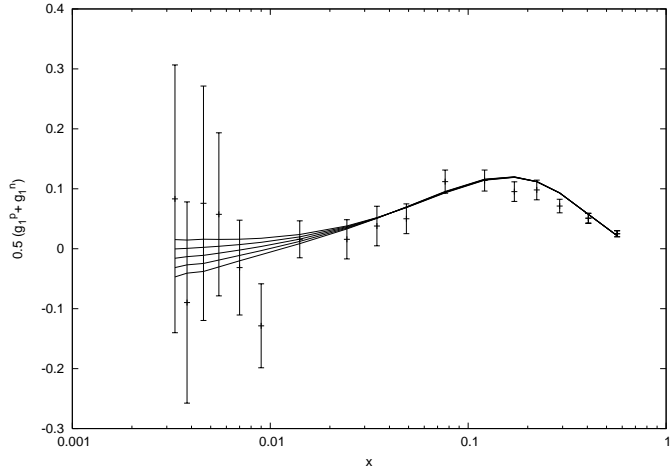}
\caption{$g_1^N = (g_1^p + g_1^n)/2$ vs $x$,
compared with COMPASS data. $Q^2$ is the measured mean value
$\langle Q^2 \rangle$ at each $x$. Results shown for five different
contributions of gluons $\Delta G$ to the proton's spin: (from up to bottom)
-0.25, 0, 0.25, 0.5, 0.75.
Error bars represent the statistical uncertainties.}
\end{center}
\end{figure}
\begin{figure}[ht]
\begin{center}
\includegraphics[width=90mm]{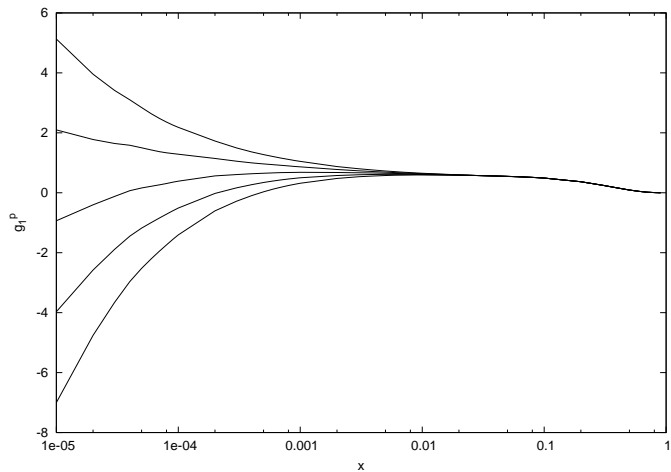}
\caption{The polarised structure function $g_1$ for the proton vs $x$ at
$Q^2=10$ ${\rm GeV}^2$. Results shown for five different
contributions of gluons $\Delta G$ to the proton's spin: (from up to bottom)
-0.25, 0, 0.25, 0.5, 0.75.}
\end{center}
\end{figure}
\begin{figure}[ht]
\begin{center}
\includegraphics[width=90mm]{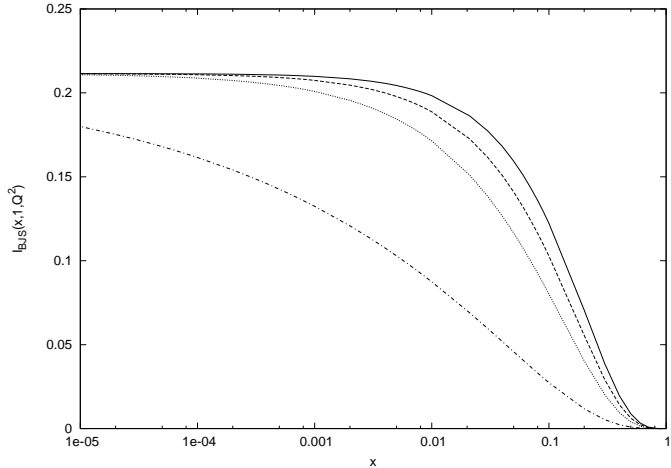}
\caption{Integral of the spin dependent nonsinglet structure
function $g_1^{NS}=g_1^{p}-g_1^{n}$ over the range $10^{-5}\leq x\leq 1$ as a function
of the low-$x$ limit of integration. $Q^2=10$ ${\rm GeV}^2$.
The comparison for different $a_1$ in the input parametrisation
$\sim x^{-a_1}(1-x)^3$ at $Q_0^2=1$ ${\rm GeV}^2$. Plots (from up to bottom):
$a_1 = 0$, 0.2, 0.4, 0.8.}
\end{center}
\end{figure}
\begin{table}[ht]
\begin{center}   
\begin{tabular}{|c|c|c|c|c|c|}\hline\hline
 & $Q^2$ & $x_1$ & $x_2$ & $\int\limits_{x_1}^{x_2}dx\: g_1$ & Experiment \\
 \hline\hline
 & & & & $0.04904^{1}$ &0.051 C \\
N & 10 & 0.004 & 0.7 &  & $\pm$0.003 (stat.) \\
 & & & & $0.04871^{2}$ & $\pm$ 0.005 (syst.) \\ \hline
 & & & & $0.1766^{a}$ & 0.1479 H \\ 
 & & & & & $\pm$0.0055 (stat.) \\
NS & 5 & 0.021 & 0.9 & $0.1718^{b}$ & $\pm$ 0.0142 (syst.) \\
 & & & & & $\pm$0.0055 (par.) \\
 & & & & $0.1486^{c}$& $\pm$ 0.0049 (evol.) \\ \hline\hline
\end{tabular}
\caption{Comparison of integrals of $g_1^N=(g_1^p+g_1^n)/2$ and $g_1^{NS}$
with COMPASS (C) and HERMES (H) data.
$g_1^N$ results for both gluon scenarios: $^{1}\Delta G<0$ and
$^{2}\Delta G>0$ are shown. For $g_1^{NS}$ the result incorporating $ln^2$
resummation $^a$ is compared with LO DGLAP solutions for singular input
parametrisations $\sim x^{-a_1}(1-x)^3$ with $a_1:$
$^{b}0.1\div 0.2$, $^{c}0.4$.}
\end{center}
\end{table}
Figs. 5-8 and Table I contain our numerical results.
Fig. 5 shows the spin dependent structure function for proton $g_1^p$
\begin{equation}\label{r5.5}
g_1^p(x,Q_0^2) = \frac{1}{2}\langle e^2\rangle\big[\;\Delta q^{S}(x,Q^2) + 
\Delta q^{NS}(x,Q^2)\;\big]
\end{equation}
as a function of $x$, compared with HERMES data \cite{b28}. Here,
$\langle e^2\rangle$ is given by (\ref{r4.0a}). We obtain our results solving
the unified evolution equations (\ref{r2.3}),(\ref{r2.4})
with `flat' parametrisations $\sim (1-x)^{a_2}$ of the parton densities.
We present plots for different values of the parameter $a_2$: 3, 2, 1 and for
a negative and positive parametrisation of gluons. In Fig. 6 we plot $g_1^N$
\begin{equation}\label{r5.6}
g_1^N = \frac{1}{2}\;(g_1^p + g_1^n) = \frac{g_1^d}{1-\frac{3}{2}\omega_D}
\end{equation}
as a function of $x$ together with COMPASS data \cite{b29}. Here, $g_1^p$,
$g_1^n$ and $g_1^d$ denotes the polarised structure function of proton,
neutron and deuteron, respectively and $\omega_D\approx 0.05$ is the D-state
admixture to the deuteron wave function.
Results are shown for different contributions of gluons to the proton's spin
at initial scale $Q_0^2$, namely $\Delta G(Q_0^2)=$ -0.25, 0, 0.25, 0.5, 0.75,
where
\begin{equation}\label{r5.7}
\Delta G(Q^2) = \int\limits_{0}^{1} dx\:\Delta G(x,Q^2).
\end{equation}
In Table I we collect the integrals of $g_1^N$ and $g_1^{NS}$ over the range
of x from COMPASS and HERMES experiments.

Our results for the function $g_1^N$ as well as its first moment are in a
very good agreement with the experimental COMPASS data.
There is certain discrepancy between our predictions and HERMES data.
This is particularly visible for the contribution to the Bjorken sum rule
\begin{equation}\label{r5.8}
I_{BJS} (x_1,x_2,Q^2) = \int\limits_{x_1}^{x_2} dx\: g_1^{NS}(x,Q^2)
\int\limits_{x_1}^{x_2} dx\: [\;g_1^p(x,Q^2)-g_1^n(x,Q^2)\;]
\end{equation}
shown in TABLE I. Some ansatz (input parametrisation) must be adopted in
two degrees of evolution. Values of $g_1$ measurements in the two or three
$Q^2$ bins for each $x$ must be evolved to their mean $Q^2$ and then averaged.
Also, the evaluation of the first moment of the structure function $g_1$
requires the evolution of all measurements to a common $Q^2$. In HERMES
analysis this is done by using a fitted parametrisation \cite{b5}, which
increases as $x\rightarrow 0$: $g_1^{NS}\sim x^{-0.8}$.
COMPASS Group have used several fits \cite{b5,b6,b30} which have been averaged.
The discrepancy between our results and HERMES data reflects the
fact that the fit used by HERMES collaboration is significantly different
from ours (\ref{r5.4}) with $a_1=0.0$. Note also that very close to the
HERMES value for $I_{BJS}$ is our result obtained within LO DGLAP approach
with use of the singular input parametrisation $g_1^{NS}(Q_0^2)\sim x^{-0.4}$.
This makes the contribution from the small-$x$ region $0<x<0.021$ more
significant - at level of $30\%$ of the total BJS compared to our estimation
based on the unified DGLAP$+ln^2(x)$ theoretical analysis, which gives about
$17\%$.
Furthermore, it can be seen from Table I that LO DGLAP evolution with the
appropriate input $\sim x^{-a_1}$ ($a_1>0$) for a given region of $x$ can
reproduce the result of the DGLAP$+ln^2(x)$ approach. In this way, a suitably
chosen initial parton density can compensate missing low-$x$ effects in
QCD analysis.

We would like also to pay special attention to the evolution
equation for truncated moments of the parton distributions. Fig. 8
illustrates the truncated contribution to the Bjorken sum as a function
of the truncation point $x$. Solving the equation for moments (\ref{r5.1}),
(\ref{r5.2}) we test different input parametrisations and find the small-$x$
contribution $I_{BJS}(0,0.01,10)$ (\ref{r5.8}) being between about $6\%$
for the flat input $\sim (1-x)^3$ and about $60\%$ for the very steep one
$\sim x^{-0.8}(1-x)^3$. The problem of the low-$x$ part of the Bjorken sum
we have also discussed in \cite{b22,b31}.

Finally, let us discuss the dependence of the polarised nucleon structure
functions on the gluon distribution $\Delta g$. From Figs. 5-6 and
Table I one can see that the predictions for $g_1^p$ and $g_1^N\sim g_1^d$
(\ref{r5.6})
in the available experimentally $x$-region ($x>0.003$) are compatible with
the data independently of the assumed gluon function. Large experimental
uncertainties for low $x$ do not allow one to discriminate between
different, in particular positive and negative polarised gluon densities.
In Fig. 7 we compare the proton structure function $g_1^p$ at
$Q^2=10$ ${\rm GeV}$ for different fractions of the nucleon spin carried by
gluons at the initial scale $Q_0^2$. Note, that $g_1^p$ essentially depends
on the gluon distribution only for very low $x$ - not before $x\approx 0.01$.
Our parametrisations of $\Delta G$ (\ref{r5.7}) reflect the latest
experimental determinations of the gluon polarisation at COMPASS \cite{b32},
RHIC \cite{b33} and STAR \cite{b34}. The shape of $\Delta G(x,Q^2)$ is poorly
known and the present experimental data support both positive and negative
distributions, resulting in small $\mid \Delta G \mid \approx$
0.2 to 0.3 (COMPASS) or large $\Delta G=-0.56\pm 2.16$ (RHIC),
$\Delta G=-0.45$ to 0.7 (STAR).
It is possible that a significant contribution to $\Delta G$ comes from
low-$x$.
A knowledge of the small-$x$ behaviour of $\Delta G(x,Q^2)$ would provide
a constraint on the shape and the sign of the gluon component. We hope future
measurements at RHIC over a wide range of $x$ and $Q^2$ will enable precise
determination of the gluon contribution to the nucleon spin.

\section{Conclusions}
\label{Concl}

In this paper we have presented results for the spin structure function $g_1$
of the nucleon together with comparison with latest HERMES and COMPASS data.
We have applied an approach that combines LO DGLAP $Q^2$ evolution with the
resummation of the double logarithmic terms $ln^2(x)$. This unified framework
goes beyond the standard LO and NLO PQCD evolution of the parton   
densities and becomes essential for $x\lesssim 0.01$. In our analysis, we have
focused on the taking into account the `very running' coupling effects. For the
small-$x$ region the more justified is the use of $\alpha_s=\alpha_s(Q^2/z)$
instead of $\alpha_s=\alpha_s(Q^2)$, with $z$ being the longitudinal momentum
fraction of a parent parton, carried by a next generation. In this way
$\alpha_s$ becomes `very running' i.e. runs in each ladder rung depending on
the gluon virtuality. We have shown, that the impact of these running coupling
effects becomes important at $x\lesssim 0.01$ and significantly damp the
results. The decreasing factor at very small $x=10^{-5}$ can be about 1/2 for
the nonsinglet and 1/6 for the singlet (gluon) distribution function in
comparison to the standard $\alpha_s(Q^2)$ prescription.

In order to calculate the first moment of $g_1$ over the available
experimentally $x$ region, we have solved the direct evolution equations for
truncated moments of the parton densities. In this approach we have utilized
the fact that the resummation of the double logarithmic terms, missing in the
standard DGLAP approximation, can be mimiced by singular input
parametrisation of the parton distributions.
The truncated moments approach refers to the physical values - moments
(rather than to the parton distributions), what in future analyses could
enable one to use a wide range of deep-inelastic scattering data in terms of
smaller number of parameters. In this way, no assumptions on the shape of
parton distributions are needed.

Our theoretical predictions for the polarised structure function $g_1$ and
its first moment for the deuteron are in a very good agreement with COMPASS
data. There is certain discrepancy between our predictions and HERMES data,
particularly visible for the contribution to the Bjorken sum. This reflects
the fact that the fit used by HERMES collaboration is significantly different
from ours. It must be emphasized, that the final (after evolution) results
strongly depend on the the input parton distributions assumed.

Finally, let us discuss the dependence of the polarised nucleon structure
functions on the gluon distribution $\Delta g$. Large experimental  
uncertainties in the low-$x$ region do not allow one to discriminate between
different polarised gluon densities. The shape of $\Delta G(x,Q^2)$ is poorly
known and the present experimental data support both positive and negative
gluon distributions. A knowledge of the small-$x$ behaviour of the gluon
component and possibly a significant contribution to $\Delta G$ from this
region would enable to resolve the nucleon spin puzzle. This is a challenge
for future theoretical and experimental efforts.

\section*{Acknowledgements}

D.K. is grateful to Boris Ermolaev for detailed explanations of the running
coupling constant problem and for interesting and inspiring discussions.\\
We would like to thank Beata Ziaja for the numerical code that was adapted
for our studies.

\end{document}